\documentclass[twoside,11pt]{article}
\usepackage{graphicx}
\usepackage{bm}
\usepackage{booktabs}
\usepackage[round]{natbib}
\usepackage[margin=1in]{geometry}
\usepackage[usenames]{color}
\usepackage{float}
\usepackage{amsfonts, amssymb, amsmath, amsthm, multicol}
\usepackage[colorlinks=true, pdfstartview=FitV, linkcolor=blue,
citecolor=blue, urlcolor=blue]{hyperref}
\usepackage[usenames]{color}
\definecolor{Red}{rgb}{1,0.05,0}
\definecolor{Grn}{rgb}{0.1,0.7,0.1}
\definecolor{Blu}{rgb}{0.1,0.1,0.6}
\definecolor{Org}{rgb}{1,0.45,0}
\definecolor{Vio}{rgb}{0.6578,0,0.9478}
\definecolor{Mag}{rgb}{1,0.2,0.3}
\usepackage{authblk}
 \usepackage{booktabs}

\title{On the Jets Induce by a Cavitation Bubble Near a Cylinder}

\author[1,2]{ {{\color{Grn}Yuxin Gou}}}
\author[1]{{\color{Org} {Junrong Zhang}}}
\author[3]{{\color{Vio} Akihito Kiyama}}
\author[1]{{\color{Red} Zhao Pan}\thanks{To whom correspondence may be addressed: zhao.pan@uwaterloo.ca}}
\affil[1]{University of Waterloo, Department of Mechanical and Mechatronics Engineering, Waterloo, ON, Canada}
\affil[2]{Department of Mechanical And Electrical Engineering, Harbin Engineering University, 
Harbin, 150001, China. }
\affil[3]{Graduate School of Science and Engineering, Department of Mechanical Science,  Saitama University, 
255 Shimo-Okubo, Sakura-ku, Saitama, 338-8570, Japan.}

\date{\today}

\begin{document}

\maketitle \vspace{ 0 cm}

\begin{abstract}  

The dynamics of cavitation bubbles in the vicinity of a solid cylinder or fibre are seen in water treatment, demolition and/or cleaning of composite materials, as well as bio-medical scenarios such as ultrasound-induced bubbles near the tubular structures in the body.
When the bubble collapses near the surface, violent fluid jets may be generated. 
Understanding whether these jets occur and predicting their directions---departing or approaching the solid surface---is crucial for assessing their potential impact on the solid phase.
However, the criteria for classifying the onset and directions of the jets created by cavitation near a curved surface of a cylinder have not been established. 
In this research, we present models to predict the occurrence and directions of the jet in such scenarios. 
The onset criteria and the direction(s) of the jets are dictated by the bubble stand-off distance and the cylinder diameter. 
Our models are validated by comprehensive experiments. 
The results not only predict the jetting behaviour but can serve as guidelines for designing and controlling the jets when a cavitation bubble collapses near a cylinder, whether for protective or destructive purposes.

\end{abstract}

\section{Introduction}\label{intro} 

Cavitation is a phase transition process from liquid to gas, which is often observed when the pressure of the liquid experiences a significant drop within a short time.
The collapse and rebound of the bubble may generate shock waves, extreme heating, and high-velocity jets, resulting in damage to the solid boundaries nearby.
This process is detrimental in many scenarios, such as cavitation erosion to hydraulic machinery and destruction of human tissues (e.g., bone or brain,~\cite{canchi2017controlled,zhang2022dynamics}).
On the other hand, some applications such as biomedical ultrasound and ultrasonic cavitation cleaning~\citep{lamminen2004mechanisms,bang2010applications} take advantage of the force acting on the boundary. 
Hence, the cavitation dynamics near the boundaries have been of interest to the community.

\begin{figure}[tbh]
    \centering
    \includegraphics
    [width=4in]{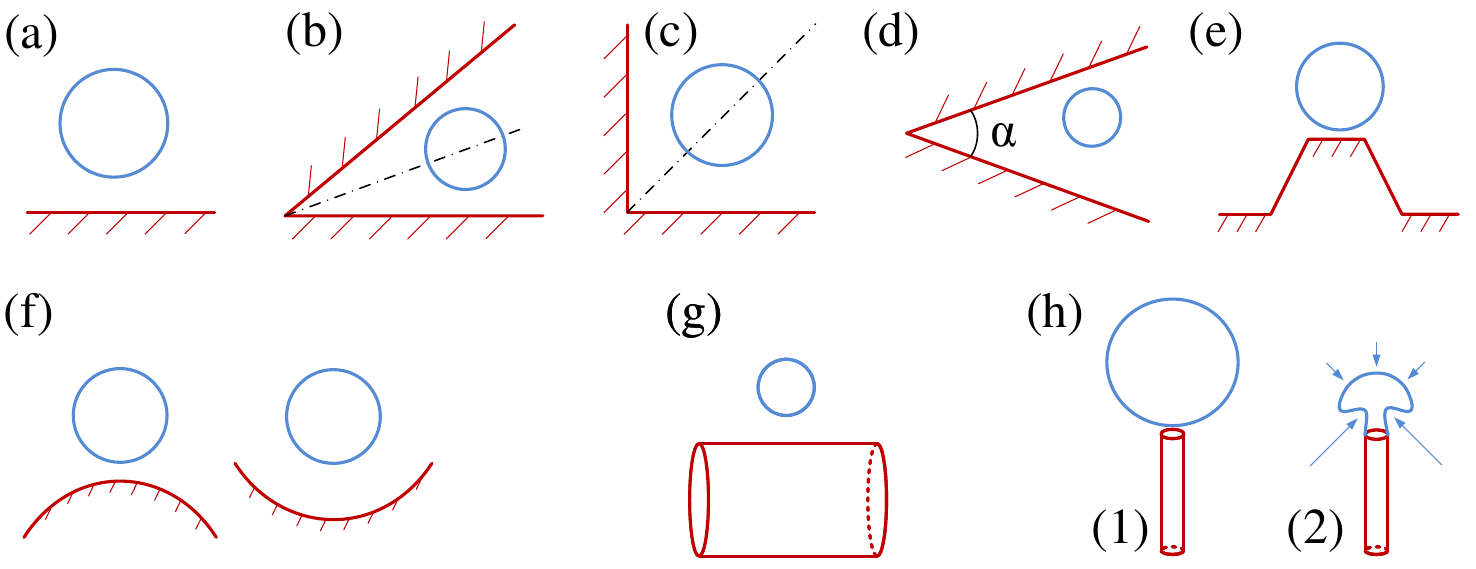}
    \caption{Schematic diagrams of previous research on a cavitation bubble (indicated by blue circles) near solid boundaries (marked in scarlet) and example studies (a)~\cite{lauterborn1975experimental},(b)~\cite{brujan2018dynamics}, (c)~\cite{cui2020experimental}, 
    (d)~\cite{tagawa2018bubble}, 
    (e)~\cite{kim2020underwater}, 
    (f)~\cite{tomita2002growth}, 
    (g)~\cite{zhang2017dynamic,mur2023microbubble}, 
    (h)~\cite{fursenko2020mechanism}.}

    \label{fig:structure}
\end{figure}

Studies on bubble dynamics near a wall and associated damaging mechanisms can be traced back to 1940's~\citep{kornfeld1944destructive}, focusing on the cavitation phenomena near a flat surface (see, for example, \citet{benjamin1966collapse,plesset1971collapse,lauterborn1975experimental,blake1986transient,supponen2016scaling}, and an illustration in figure.~\ref{fig:structure}(a)).
When a bubble collapses near a flat solid wall, the bubble may migrate to the wall, and a directional liquid jet towards the wall is created.

The concentrated momentum impacts a small area on the wall, where the induced pressure and shear are considered to be one of the primary mechanisms for cleaning and/or damaging the surfaces~\citep{dular2004relationship,wang2015bubble,supponen2016scaling,gonzalez2021acoustic}.
Therefore, the onset of the directional jet is the key factor determining the interaction between the bubble and the boundary.

The direction of the jet depends on a multitude of factors, especially the geometry of the boundaries. 
\cite{wang2014experimental,brujan2018dynamics,cui2020experimental} experimentally studied the direction of the jet generated upon the rebound of a bubble in a corner of two solid boundaries, where the angle between them was set to either 90~$\deg$ or less   (figure.~\ref{fig:structure}(b~\&~c)).
\cite{tagawa2018bubble} proposed a generalized formula that predicts the jet direction in a corner with an arbitrary opening angle $\alpha$ and proximity to the walls (figure.~\ref{fig:structure}(d)). 
They show that there exist analytic solutions that predict the jet direction for $\alpha = \pi/n$, where $n$ is a natural number.

Several studies reported that the fluid jet formed upon the bubble collapse near a solid wall with complex geometry does not always point to the wall. 
\cite{kim2020underwater} reported the dynamics of the bubbles near trapezoidal ridges and valleys (figure.~\ref{fig:structure}(e)) and found that the fluid jet can appear in two different directions (i.e., a departing or approaching jet to the wall).
The departing jet may appear when a bubble collapses near the ridge, while a bubble near the valley can only form an approaching jet in their experiments. 
The configuration might share some similarity to the bubble dynamic near a curved surface (e.g., the surface of a cylinder or a sphere, see figure~\ref{fig:structure}(f~\&~g)).
The morphology of the bubble in the neighbourhood of a curved surface has been studied \citep{tomita2002growth,zhang2013experimental}, and the curvature of the solid wall was found to be one of the primary parameters in addition to the stand-off distance \citep{takahira1989collapse}.
A departing jet may appear when the bubble collapses near a convex (positive curvature) surface.
However, extensive data or detailed discussions on the direction of the dual fluid jets were not reported.

An interesting feature of the bubble near a convex surface is the ``mushroom'' bubble before collapsing, which is almost always associated with the departing jet.
This observation has been reported in earlier studies (e.g., \cite{tomita2002growth,zhang2013experimental}) and recent research on cavitation near the tip of a thin cylinder also concurred with similar evidence. 
\cite{fursenko2020mechanism,koch2021dynamics} reported that the mushroom-shaped collapsing bubble could happen when a cavitation bubble was initiated near the tip of a thin cylinder (figure.~\ref{fig:structure}(h)). 
The fluid-gas interface resembling the `stem of the mushroom' (i.e., the interfaces close to the tip of the cylinder) contracts faster than the `mushroom cap', which results in a departing jet when the bubble fully collapses.
\cite{fursenko2020mechanism} also suggested that an optimal length scale of cylinder thickness exists, compared to a fixed bubble diameter, so that the jet becomes the most powerful.
\cite{kadivar2021dynamics} numerically approached this problem and revealed that the mushroom-shaped bubble near the tip of the cylinder might be linked to the reduction of the impact load on the surface. 
It is perhaps because the not-yet-formed departing jet carries momentum away from the solid surface.
Beyond the distinct physics, this setup of bubbles near the tip of a thin cylinder can generate a high-speed departing jet (up to $O(1000)$~m/s according to the simulations by \cite{koch2021dynamics}) and is of interest to applied research. 
However, the direction of the jets and the criteria of the departing jet onset were not analyzed.

In the current work, we are interested in the dynamics of bubbles and jets next to the side surface of cylinders. 
To the best of our knowledge, this scenario has not been reported except for \cite{mur2023microbubble} studying the micro-bubbles near a fiber, as well as \cite{zhang2013experimental} where bubble behaviour near a thick cylinder (inspired by cavitation near the hull of a ship) was investigated.
There are no detailed discussions on the direction of the jet(s) when the bubble collapses near a cylinder available in the current literature. 
In this paper, we report a regime diagram, validated by vast experimental data, that classifies the onset and the direction of the jet(s), which is dictated by two non-dimensional parameters (i.e., bubble stand-off distance and the cylinder thickness relative to the bubble diameter). 
Particularly, we find that when a large bubble is close to a thin cylinder, a departing jet is likely to form after collapsing and the cylinder is protected. 
This discovery might be insightful for some applied scenarios. 
For example, fibrous or tubular structures in the vicinity of a cavitation bubble could be free from severe damage and it is possible to design patterned surface \citep{kim2020underwater} or fibrous structure to reduce cavitation erosion.

\section{Experimental setup} 
\label{setUpSec}

\begin{figure}[tbh]
    \centering
    \includegraphics[width=0.9\textwidth]{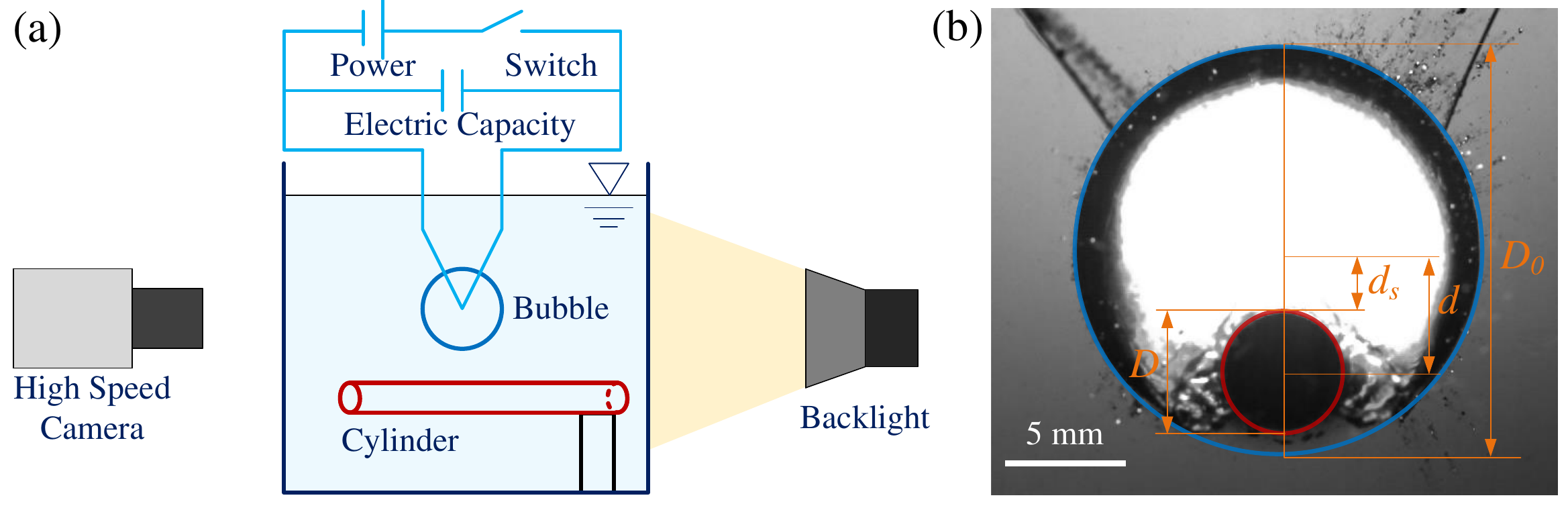}
    \caption{Schematic diagram of the experimental setup (a) and the dimensions of the bubble and cylinder marked on a high-speed image (b). (a) is not to scale for illustration.} 
    \label{fig_setup}
\end{figure}

The experimental setup is shown in figure~\ref{fig_setup}(a). 
The cavitation bubbles were generated by shorting adjustable direct current voltage carried by two thin wires of 0.14~mm in diameter.
The sizes of the bubbles varied from 5.45 to 24.58~mm in diameter by adjusting the voltage (within the range of 60~--~120~V).  
The cylinders used in the experiments are made from stainless steel with a contact angle of around $60^\circ$.
The wires are at least one order of magnitude thinner compared to the size of the cylinders and the cavitation bubbles and thus the influence of the wires is negligible.
The wires and the cylinder were placed in the middle of a tank ($20 \times 20 \times 20$~cm$^3$) filled with degassed tap water. 
The tank is large enough to ensure the bubble behaviour was not affected by either the free surface or the rigid wall.
The dynamics of the cavitation bubbles was filmed by a high-speed camera (FASTCAM SA-Z or NOVA S20, Photron, Tokyo, Japan) at 60,000 frames per second.

A schematic of the bubble and the cylinder overlaid on a high-speed image is shown in figure~\ref{fig_setup}(b). 
Two key non-dimensional parameters---the standoff distance $\gamma$ and the non-dimensional cylinder diameter $\eta$---are  defined as
\begin{equation}
\label{eq:non-dimensional}
     \gamma =\frac{d_s}{D_0} \; ~\text{and}~ \;  \eta = \frac{D}{D_0},
\end{equation}
respectively, where $d_s$ is the distance from the spark location, which can be considered as the nominal center of the bubble, to the closest cylinder surface, $D_0$ is the maximum bubble diameter (marked by a blue circle), and $D$ is the cylinder diameter (marked by a red circle). 
The distance between the nominal center of the bubble and the center line of the cylinder is written as $d = d_s + D/2$, which can be normalized by $D_0$ as 
\begin{equation}
	\label{eq: zeta}
	\zeta = \frac{d}{D_0} = \gamma +\frac{\eta}{2}.
\end{equation}
This is an alternative non-dimensional length scale characterizing the distance between the bubble and the cylinder.

\section{Results}
\label{results}
\begin{figure}[!htb]
    \centering
    \includegraphics[width=\textwidth]{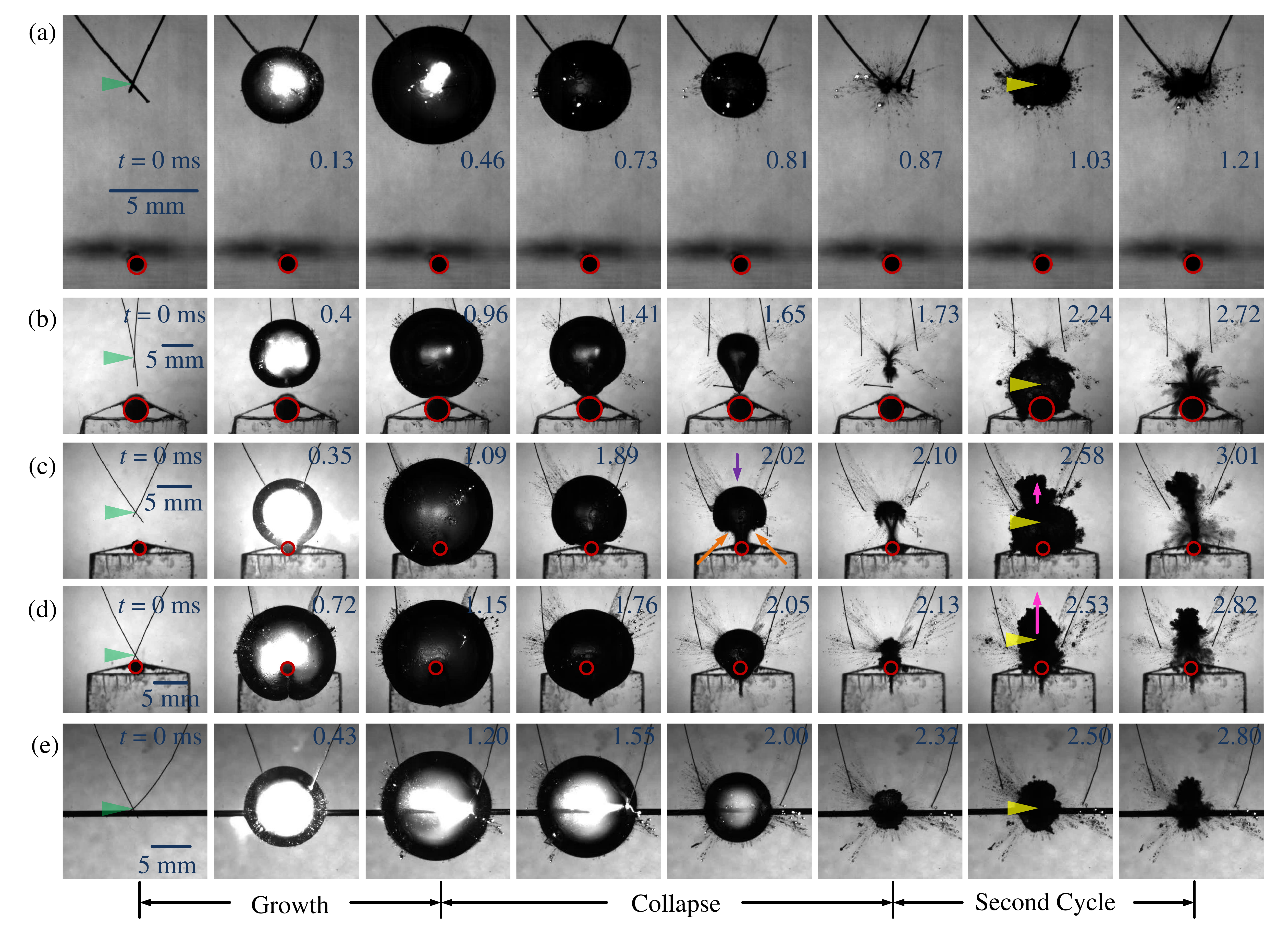}
    \caption{High-speed images of five characteristic behaviours of a cavitation bubble near a cylinder observed for various $\gamma$ and $\eta$. (a)~A no jet case with a small value of $\eta$, (b)~approaching jet case, (c)~approaching jet dominant and departing jet emerging case, (d)~departing jet dominant, and (e)~a no jet case due to large $\gamma$, viewed from the side. (a~--~d) are viewed from the front. The green and yellow triangles mark the location of the spark and the centroid of the bubble cloud, respectively. The coloured arrowheads qualitatively illustrate the direction and the amplitude of the liquid jet or the moving bubble cloud. The scarlet circles mark the locations of the cylinders.}
    \label{fig_results}
\end{figure}

We carried out comprehensive experiments on spark-induced cavitation bubbles in the vicinity of a cylinder by varying $\eta$ and $\gamma$.
The experiments revealed five distinct bubble behaviours for various conditions (demonstrated in figure~\ref{fig_results}). 
The dimensional and non-dimensional parameters of these typical cases are listed in~Table~\ref{tab:parameters}.

When the bubble is initiated far enough from the surface of a cylinder, it is expected that the bubble remains spherical when expanding and collapsing, and no jets are formed after the bubble collapses. 
We refer to this observation as a ``no jet (NJ)'' case hereafter.
For example, in figure~\ref{fig_results}(a), a bubble is initiated by a spark (indicated by the apex of the green triangle at $t=0$~ms) at $\gamma=1.44$ from a cylinder (marked by the scarlet circle).
The bubble grows and reaches its maximum diameter $D_0$ at $t = 0.46$~ms, collapses at $t = 0.87$~ms for the first time, and rebounds to the maximum of the cloud at $t=1.03$~ms. 
The direct observation of the jets (onset and directions) during collapse can be difficult, thus we use the displacement ($\delta_D$) from the bubble onset location (marked by the green triangle in figure~\ref{fig_results}) to the centroid of the maximum bubble cloud of the second expansion (marked by the yellow triangle) as an indicator of the net momentum due to the bubble collapse. 
The positive direction of $\delta_D$ points from the centerline of the cylinder to the center of the bubble). 
A non-zero $\delta_D$ infers a liquid jet generated when the bubble collapses. 
The non-dimensional displacement, $\delta = \delta_D/D_0$, in figure~\ref{fig_results}(a) was $\delta = 0.00$ (note that NJ is classified for $|\delta|<\delta_0,~\delta_0 = 0.03$ is a small value as the measurement threshold in this work.)

\begin{table}[h]
  \begin{center}
    \caption{Dimensional and non-dimensional parameters of the cases shown in figure~\ref{fig_results}.}
  \label{tab:parameters}
\def~{\hphantom{0}}
  \begin{tabular}{lccccccccc}
  \toprule
Cases & Voltage (V) & $D_0$ (mm) & $D$ (mm) & $d$ (mm) & $d_s$ (mm) & $\delta_D$ (mm) & $\gamma$ & $\eta$ & $\delta$ \\ \midrule
(a)   & 60          & 7.71         & 1.00       & 11.61       & 11.11         &\;\; 0.00             & 1.44       & 0.13     &\;\; 0.00         \\
(b)   & 120          & 16.41          & 4.00       & 9.39       & 7.39         & $-2.01$              & 0.45       & 0.24     &  $-0.12$         \\
(c)   & 120          & 20.45         & 2.00       & 6.41       & 5.42         & $-1.02$              & 0.26       & 0.10     & $-0.05$         \\
(d)   & 120          & 20.61         & 2.00       & 2.25       & 1.25         &  +0.75              & 0.06       & 0.10     & +0.04         \\
(e)   & 120          & 19.95         & 1.00       & 0.83       & 0.33         & \;~0.00             & 0.02       & 0.05     &\;~0.00         \\ 
   \bottomrule
  \end{tabular}
      \end{center}
\end{table}

As the center of the bubble moves closer to the cylinder, a jet shooting toward the cylinder is generated when the bubble collapses and we address this case as ``approaching jet only (AJO)". 
As shown in figure~\ref{fig_results}(b) as an example, the bottom of the bubble is deformed when approaching the cylinder from a standoff distance of $\gamma = 0.45$ (e.g., see two frames at $t = $~0.40~and~0.96~ms).
The centroid of the rebound bubble (marked by the yellow triangle at $t = 2.24$~ms)  moves towards the cylinder ($\delta=-0.12$ in this case), compared to the spark location (marked by the green triangle at $t = 0$~ms).
This footprint indicates a liquid jet approaching the cylinder is generated during the bubble rebound. 
In addition, no other jet(s) were observed. 
The bubble cloud formed during the second expansion cycle collapses and largely covers the cylinder ($t = 2.72$~ms), implying that the approaching jet may carry a large momentum. 
This process that generates an approaching jet is similar to a bubble collapsing near a flat rigid surface.

Figure~\ref{fig_results}(c) presents a typical case where the mushroom bubble forms and a departing jet starts to appear. 
In this work, we refer to this scenario as ``departing jet emerging (DJE)''. 
The stand-off distance $\gamma = 0.26$ and the non-dimensional cylinder size $\eta=0.09$ in this case were smaller than those of the case in figure~\ref{fig_results}(b).
In figure~\ref{fig_results}(c), when the bubble reaches its maximum volume (at $t = 1.09$~ms), the bubble partially warps the narrow cylinder and maintains its spherical shape in general.
The stem of the ``mushroom" is formed due to the fast-retracting liquid jets pinching the bubble near the cylinder (indicated by the orange arrowheads at $t =$~2.02~ms). 
While collapsing, the cap of the mushroom remains spherical as the gas-liquid interface (indicated by the purple arrowhead at $t =2.02$~ms) is far away from the cylinder and recedes slower compared to the pinching jets. 
The dynamics are similar to the observations made by \cite{zhang2013experimental,fursenko2020mechanism,koch2021dynamics}.  

It is noteworthy that the bubble cloud in the second expansion cycle moves in two directions.
The centroid of the rebound bubble moves toward the cylinder ($\delta= -0.05$, comparing the location of the green and yellow triangles at $t = ~0~\text{and}~2.58$~ms, respectively), similar to the case in figure~\ref{fig_results}(b), while there is a minor cloud bubble shooting away from the
cylinder (see $t = 2.58$~ms, marked by the short pink arrowhead in figure~\ref{fig_results}(c)). 
This observation indicates that two jets exist after the collapse: one jet is approaching and the other one is departing from the cylinder. 
The departing jet, which is an emerging  feature compared to the case in figure \ref{fig_results}(b), however, does not yet dominate the entire jetting process.


 
When the bubble is close to a relatively thin cylinder, the departing jet may dominate over the approaching jet and we denote this scenario as ``departing jet dominant (DJD)''.
A typical case is shown in figure~\ref{fig_results}(d) for $\gamma = 0.06$ and $\eta = 0.09$. 
The bubble completely wraps the cylinder when it expands to the maximum diameter ($t= 1.15$~ms) and then collapses. 
Similar to the case shown in figure~\ref{fig_results}(c), the elongated rebound bubble cloud covering the cylinder meanwhile moving away from the ($t=2.53$~ms) indicates the existence of both approaching and departing jets.
Noting centroid of the bubble cloud ($t= 2.53$~ms, marked by the yellow triangle) is further away from the cylinder than the center of bubble onset (green triangle at $t= 0$~ms) and the corresponding displacement $\delta = +0.04$, we argue that the jet forming at collapse is mainly departing.

Figure~\ref{fig_results}(e) shows another ``no-jet (NJ)'' case. 
A bubble is initiated right next to a thin cylinder, where the size of the bubble is much larger than that of the cylinder ($\eta = 0.05$). 
The bubble behaviour in this case is similar to a free bubble.
The centroid of the bubble (cloud) does not show any apparent movement upon rebound, indicating that no jet was generated. 
Despite the NJ outcome that is similar to the case shown in figure \ref{fig_results}(a), we emphasize that the phenomenon shown in figure \ref{fig_results}(e) is due to vanishing cylinder diameter ($\eta \rightarrow 0$) whereas the NJ case in figure~\ref{fig_results}(a) is associated with the standoff distance in the limit of $\gamma\rightarrow\infty$.

\section{Mechanisms}
 \label{mechanism}
The observations in figure~\ref{fig_results} imply that when a bubble collapses near a cylinder, depending on the relative position as well as the size of the bubble and the cylinder ($\gamma$ and $\eta$), the cylinder may affect the liquid flow in two ways (i.e., blocking and focusing).

First, the cylinder can \textit{block} the liquid behind it from directly moving to the center of the bubble, while the liquid on the other side of the bubble is free to move to fill the cavity during collapsing.
This causes a pressure gradient and, in turn, the collapsing bubble generates a jet approaching the cylinder~\citep{supponen2016scaling}. 
This often happens when the cylinder is relatively large and/or the bubble is not too close to the cylinder (e.g., see the case in figure~\ref{fig_results}(b)).
This mechanism is similar to the well-known jet formation from a bubble collapsing next to a solid flat surface. 

Second, when the cylinder is relatively small and the bubble is initiated close enough to the cylinder, the bubble can be significantly deformed during its growth. 
In figure~\ref{fig_results}(c), for example, the bubble partially wraps the cylinder while achieving its maximum volume (at $t = 1.09$~ms), leaving two regions of the gas-liquid interface having a higher curvature than other parts of the bubble. 
The higher curvature is corresponding to a smaller equivalent local bubble radius, which is associated with a shorter time for a local collapse.
This mechanism has also been argued by \cite{10.1007/978-3-642-51070-0_7} based on the Rayleigh's collapse time, $T \simeq 0.915\tilde{D}_0\sqrt{\rho /p_\infty}$, where $T$ is the collapse time, $\rho$ is the liquid density, $p_\infty$ is the ambient pressure, and $\tilde{D}_0$ is the equivalent bubble size reflecting the local curvature of the bubble.
Over the initial stage of the collapse, the advantage of the high-speed flows driven by the high curvature interface accumulates, which results in two jets pinching the bubble (see the orange arrowheads in figure~\ref{fig_results}(c) for instance). 
The two pinching jets forms the stem of the mushroom-shaped bubble before collapsing. 
After pinch-off, the two pinching jets merge and the momentum is \textit{focused} upward, pointing away from the cylinder, which can dominate the retracting liquid near the cap of the mushroom-shaped bubble (see the purple arrowhead in figure~\ref{fig_results}(c)). 
This focusing mechanism is similar to the shaped charge effect. 
The competition between these two mechanisms dictates the onset and direction(s) of the jet(s), and some typical results as shown in figure~\ref{fig_results}. 

\section{Regime Diagrams and Validation}
\begin{figure}[!htb]
	\centering
	\includegraphics[width=0.99\textwidth]{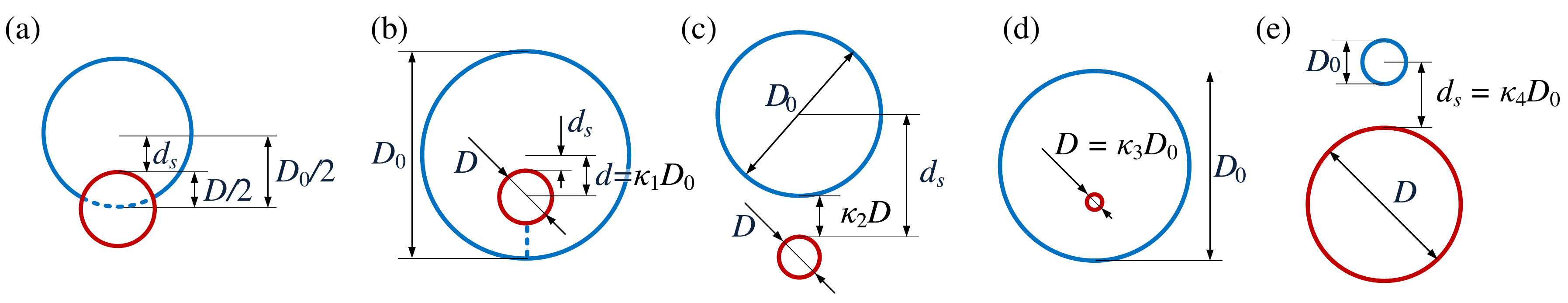}
	\caption{Schematic diagram of the critical positioning and size of bubbles (blue circles) and cylinders (indicated by scarlet circles). }
	\label{fig_distance}
\end{figure}

Based on the above experimental observations and analysis on the mechanisms, we hypothesize that the direction(s) of the jet(s) caused by the bubble collapsing near a cylinder are dictated by two parameters. 
One is the standoff distance $\gamma = d_s/D_0$ measuring the distance from the bubble to the cylinder, and the other is the non-dimensional cylinder diameter $\eta = D/D_0$. 
Several critical states regarding $\gamma$ and $\eta $ are proposed below and illustrated in figure~\ref{fig_distance}.

When a bubble wraps about half of the cylinder, the virtual circle enclosing the bubble passes the center of the cylinder (see figure~\ref{fig_distance}(a)). 
We conjecture that this is a state separating the blocking and focusing mechanisms and determines if a departing jet would emerge. 
The corresponding geometric relationship for the circles representing the bubble and cylinder is $d_s=\frac{1}{2}(D_0-D)$, and the non-dimensional form is $\gamma = \frac{1}{2} - \frac{1}{2} \eta.$ 
If the standoff distance is smaller than this threshold, that is to say 
\begin{equation}
\gamma < \frac{1}{2} - \frac{1}{2} \eta,
    \label{boundary_1}
\end{equation}
high curvature on the sufficiently deformed bubble leads to the evident focusing effect and a departing jet is expected.

When the bubble is even closer to the cylinder, especially when the bubble is relatively large, the focusing effect is more pronounced than the blocking and the departing jet starts to dominate.
This condition translates to $d < \kappa_1 D_0$, where $\kappa_1$ is a coefficient that can be determined by experimental data (see figure~\ref{fig_distance}(b) for illustration). 
Invoking $d = d_s+\frac{1}{2}D$, the non-dimensional form of this criterion is 
\begin{equation}
\gamma < \kappa_1 - \frac{1}{2}\eta.
    \label{boundary_2}
\end{equation}

When the bubble is far enough from a sufficiently small cylinder,  $d_s > \frac{1}{2}D_0 + \kappa_2 D$, where $\kappa_2$ is another constant to be determined (see figure~\ref{fig_distance}(c)), the effect of the cylinder (blocking or focusing) is negligible and thus no jet is expected. 
The corresponding non-dimensional form is 
\begin{equation}
\gamma > \frac{1}{2} + \kappa_2 \eta.
    \label{boundary_3}
\end{equation}
This criterion considers the combined effects of the relative size and position of a bubble and cylinder.

The asymptotic behaviours (i.e., small $\eta \rightarrow 0$ and large $\gamma \rightarrow \infty$) of such a setup are also of interest.
When the cylinder is significantly smaller than the bubble (see figure~\ref{fig_distance}(d) for illustration), for example, $D < \kappa_3 D_0 \ll D_0$ with corresponding non-dimensional form
\begin{equation}
\eta < \kappa_3 \ll 1,
    \label{boundary_4}
\end{equation}
the relative placement of the bubble and cylinder is not important anymore. 
Jets are not expected when the bubble collapses due to the diminishing impact of the cylinder of a small length scale. $\kappa_3 \ll 1$ is a small constant that can be found by experiments.  
When the bubble is too far away from the cylinder (see figure~\ref{fig_distance}(e)), the size of the cylinder does not matter.
We expect there exists a critical value $\kappa_4$ so that if $d_s > \kappa_4 D_0 \gg D_0/2$, no jet would be generated when the bubble collapse. 
The non-dimensional form of this criterion is 
\begin{equation}
\gamma > \kappa_4 \gg \frac{1}{2}.
    \label{boundary_5}
\end{equation}
Recall \eqref{eq: zeta} again, the above criteria can also be expressed using $\zeta$ instead of $\gamma$. 
We use $\gamma$ to be consistent with the current literature, however, $\zeta$ is practical to investigate some of the critical states regarding the directions of the jets.

The directions of the jets after bubble collapsing can be qualitatively observed by the direction of the moving bubble cloud in the high-speed videos. 
For example, when a departing jet appears, the bubble cloud tends to move away from the cylinder over the collapsing-rebound cycles.
This can be quantitatively identified using the value of $\delta = \delta_D/D_0$ as a measure, which is a characteristic displacement of the bubble cloud.
If there is only an approaching jet appears after the first collapse, the momentum of the jet would carry the bubble cloud towards the cylinder (e.g.,~see figure~\ref{fig_results}(b)) and we expect $\delta < -\delta_0<0$.
Similarly, when the departing jet dominates the approaching one, $\delta > +\delta_0>0$ (see figure~\ref{fig_results}(d) for instance).
However, if the departing and approaching jets cannot dominate one to the other, the direction of $\delta_D$ and the `sign' of $\delta$ are not necessarily determined.

\begin{figure}[!htb]
    \centering
    \includegraphics[width=0.8\textwidth]{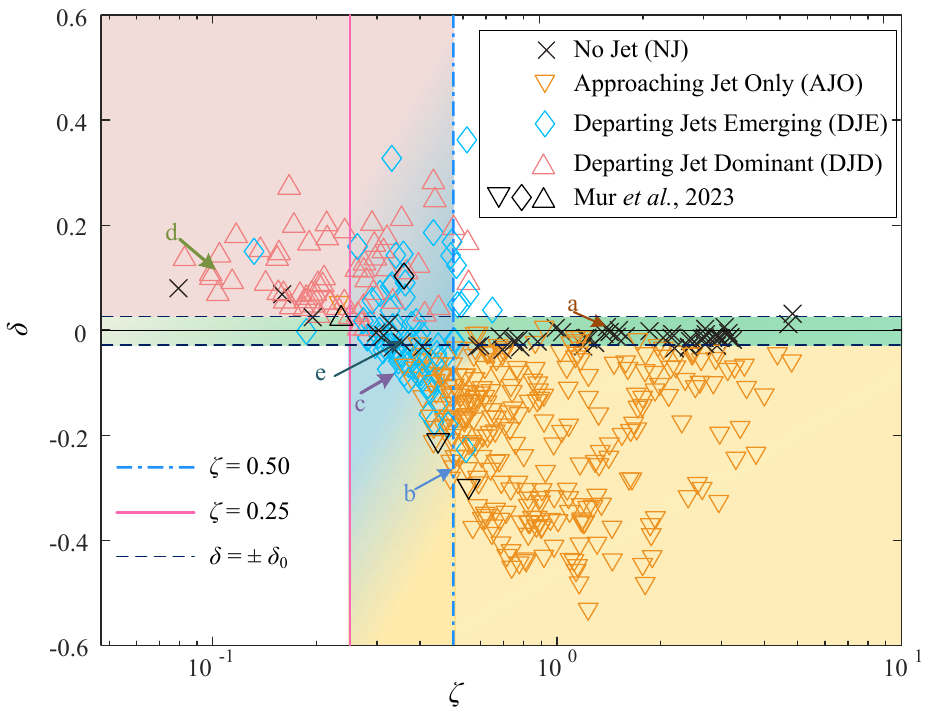}
    \caption{Experimental data (symbols with various colors and shapes) on the $\delta$--$\zeta$ phase diagram, $\delta_0=0.03$. Arrowheads (a)~--~(e) identify the data points from the experiments shown in figure~\ref{fig_results}(a)~--~(e), respectively. 
}
    \label{fig_deltad}
\end{figure}

We present $\delta$ as a function of $\zeta$ in figure~\ref{fig_deltad} to show our argument above is valid.
Viewing the $\delta$--$\zeta$ phase diagram vertically, we can see that all the AJO cases (orange upside-down triangle in figure~\ref{fig_deltad}) are located in the region of $\delta<-\delta_0$, whereas DJD cases (pink upright triangles) are in $\delta>+\delta_0$. 
NJ cases (black crosses) are distributed along $\delta = 0$ ($ -\delta_0 < \delta < +\delta_0$ to be more specific) whereas the DJE cases (blue diamond symbols) are scattered on both sides of $\delta = 0$.

Interrogating the experimental data on the $\delta$~--~$\zeta$ phase diagram ( figure~\ref{fig_deltad}) horizontally is  useful for verifying the aforementioned models and identifying the coefficients such as $\kappa_1$.
It is visible that the jet direction evolves from approaching to departing as $\zeta$ decreases.
In the region of $\zeta > 0.5$ (yellow-shaded, to the right of the blue chain line in figure~\ref{fig_deltad}), almost only AJO cases exist.
Recalling \eqref{eq: zeta}, $\zeta= 0.5$ is an alternative expression of $\gamma=1/2-1/2\eta$, thus, \eqref{boundary_1} is validated.
The departing jets emerge when $\zeta < 0.5$, 
and further reducing $\zeta$, the departing jet eventually becomes dominant for $\zeta < 0.25$, which is equivalent to \eqref{boundary_2} for $\kappa_1 = 1/4$.
This is supported by observing that in the red-shaded region to the left of the magenta line (corresponding to $\zeta = 0.25$), almost only DJD cases exist. 
The DJE cases (blue diamond symbols) are located in the transient region for $0.25<\zeta < 0.5$.

The black symbols represent the data extracted from \cite{mur2023microbubble}, where a laser-induced micro-bubble collapsing near a micro-fibre was studied. 
This work did not focus on the direction of jets, and the bubble dynamics after the first collapse was not reported. 
Instead, the location of the bubble near collapsing was recorded. 
Comparing the displacement from the location of the bubble onset to the center of the bubble at the first collapse, one could still infer the directions of jets. 
Despite being a different measure of $\delta$ than we used for our data, this qualitative classification is sufficient to tell the AJO, DJE, and DJD cases apart in \cite{mur2023microbubble}, and we see that the experimental data by \cite{mur2023microbubble} agree with our model.

\begin{figure}[!htb]
    \centering
    \includegraphics[width= 0.8\textwidth]{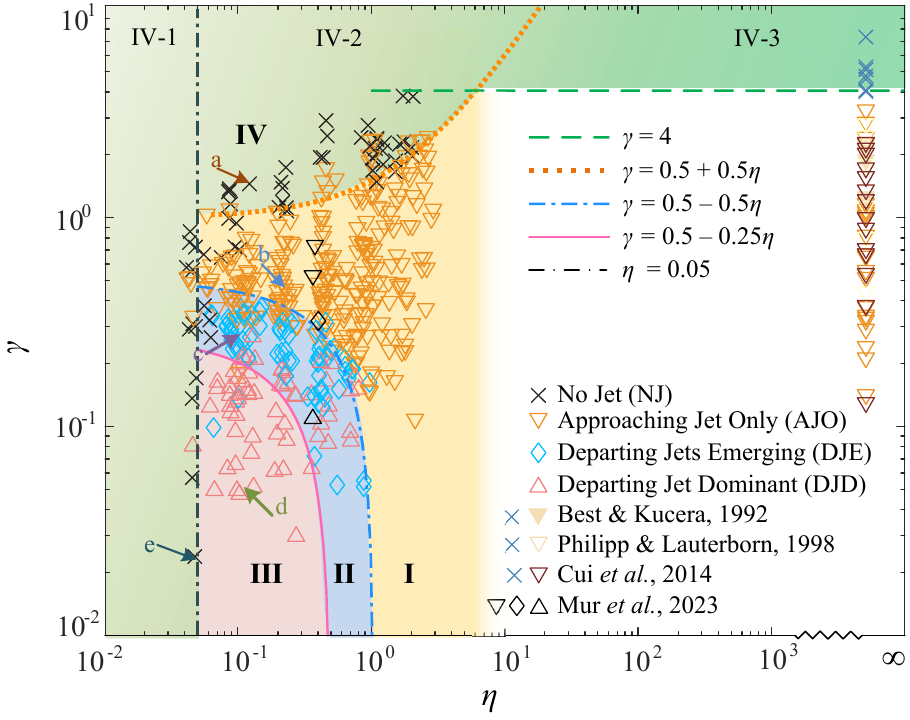}
    \caption{Regime diagram for the jet onset and directions based on cylinder size $\eta$ and standoff distance $\gamma$. Arrowheads a~--~e identify the data points from the experiments shown in figure~\ref{fig_results}(a)~--~(e), respectively. }
    \label{fig_jets direction}
\end{figure}

To validate equations \eqref{boundary_1} and \eqref{boundary_2}, we plot the non-dimensionalized experimental data on the $\gamma$~--~$\eta$ plane (figure~\ref{fig_jets direction}). 
The blue chain line indicates equation \eqref{boundary_1} separating the AJO and the DJE cases.
The magenta line in figure~\ref{fig_jets direction} is based on equation \eqref{boundary_2} that separates most DJD cases from the DJE cases. 

Experimental data on the $\gamma$~--~$\eta$ plane also provides quantitative insights into the NJ cases due to different reasons. 
$\kappa_2=0.5$ for \eqref{boundary_3} separates the NJ cases and the AJO cases for $5\times10^{-2} \lesssim \eta \lesssim 7$ (see the orange dotted line in figure~\ref{fig_jets direction}). 
For $\eta < 5\times10^{-2}$, a sufficiently thin cylinder cannot affect the dynamics of the bubble and almost no jets were observed in our experiments. 
Thus, $\kappa_3 = 5\times10^{-2}$ in \eqref{boundary_4} allows our model to establish the criterion of a thin cylinder.
For the other extreme, $\kappa_4 = 4$ for equation \eqref{boundary_5} was suggested by our experiments, which is the criterion for a large stand-off distance.
We note that $\kappa_4=4$ agrees with the established data about a cavitation bubble near a flat surface (\cite{best_kucera_1992,philipp1998cavitation,cui2013experimental}), which can be considered as a thick cylinder with vanishing curvature (i.e., $\eta \rightarrow \infty$).

In figure~\ref{fig_jets direction}, criteria based on equations~\eqref{boundary_1}~--~\eqref{boundary_4} separate the $\gamma$~--~$\eta$ phase diagram into four regimes. 
Regime I (yellow shade) covers most of the AJO cases (orange upside-down triangles). 
In regime III (pink shade), almost only pink triangles (associated with DJD cases) appear.
The transient cases for the directional jet(s) (DJE cases, marked by the blue diamond symbols in blue-shaded regime II) are in between Regimes I and III.
Regime IV (different shades of green for three sub-regimes) indicates NJ cases rooted in different mechanisms.
In Regime IV-1, NJ happens as a cylinder is too thin (small $\eta$). 
In Regime IV-3, NJ is expected as the bubble is too far away from the solid surface (large $\gamma$). 
Regime IV-2 can be thought of as the transient region between Regime IV-1 and IV-3, where the combined effect of $\eta$ and $\gamma$ must be considered and is governed by 
\eqref{boundary_1}.
Again, the data extracted from \cite{mur2023microbubble} falls in our regime diagram, and provides additional validation based on the interaction of micro-bubbles.

\section{Concluding Remarks} 
In the current work, we carried out systematic experiments to investigate a cavitation bubble collapsing near a cylinder. 
We find that the onset and the direction of the jet(s) are dictated by the relative positioning and the size of the bubble and the cylinder (i.e., the standoff distance $\gamma$ and the normalized cylinder diameter $\eta$).
When the cylinder is too thin and/or too far away from the bubble, a bubble does not expel any visible jets. 
Once the bubble starts interacting with the cylinder---when $\gamma$ and/or $\eta$ are small enough---a jet approaching the cylinder occurs, as one might expect, which is similar to that for a bubble collapsing in the vicinity of a flat wall. 
When the cavitation bubble is onset closer to an even smaller cylinder within a particular range, the bubble possesses a mushroom-like collapse followed by a departing jet.
Given a certain maximum bubble size, the departing jet carries the energy 
away from the cylinder, which might result in a reduction of the cavitation-induced damage. 
In this sense, the cylinder is protected by being thin and staying close to the cavitation.
We proposed models to classify these phenomena including transition into four regimes on the $\gamma$~--~$\eta$ phase diagram, which are validated by experiments.

The experimental results and criteria
shown in this work may be of interest to applications where cavitation bubbles interact with (thin) cylinders and fibres.
For example, a direct implication based on our result is that the demolition of thin fibres and fibrous materials could be challenging, and small bubbles are more effective than bigger ones. 
When a cylinder near a cavitation bubble needs protection, our regime diagram provides a guideline: one may want to manage the standoff distance and bubble size to avoid the jet onset or staying in the departing jet dominant regime.

\section*{Acknowledgments} 
We thank Drs. S.~Peterson and M.~Worswick for lending us equipment and J.~Beginner and J.~Imbert-Boyd for manufacturing and technical support.

\bibliography{RodLibrary}

\end{document}